\documentclass[letterpaper,10pt]{article} 
\usepackage{opticameet3} 
\newcommand\authormark[1]{\textsuperscript{#1}}
\usepackage{amsmath,amssymb}
\usepackage[colorlinks=true,bookmarks=false,citecolor=blue,urlcolor=blue]{hyperref} 
\usepackage{dsfont,color,float,empheq}
\usepackage{graphicx,wrapfig,multicol,multirow}
\usepackage{psfrag,wrapfig}
\usepackage[font=footnotesize]{caption}
\usepackage{pgfplots}
\usepackage{subcaption}
\usepackage{cite}
\usepackage{setspace}
\usepackage{multicol}
\usetikzlibrary{patterns}
\usepackage{tikz,pgfplots}
\usetikzlibrary{positioning,shadows,shapes,backgrounds,calc}
\usepackage{pgfplotstable}
\usepackage{booktabs}
\tikzstyle{arrow} =[thick,->,>=stealth] 
\usepgfplotslibrary{fillbetween}

\definecolor{my_green}{rgb}{0.25, 0.6, 0.0}    
\definecolor{my_purple}{rgb}{0.5, 0, 0.5}
\definecolor{azure}{rgb}{0, 0.5, 0.5}

\begin{document} 

\title{Nonlinearity Cancellation Based on Optimized First Order Perturbative Kernels} 

\author{Alex Alvarado\authormark{*},  Astrid Barreiro, and Gabriele Liga}

\address{Department of Electrical Engineering, Eindhoven University of Technology, Eindhoven, The Netherlands}

\email{\authormark{*}a.alvarado@tue.nl} 

\begin{abstract}
The potential offered by interference cancellation based on optimized regular perturbation kernels of the Manakov equation is studied. Theoretical gains of up to $2.5$~dB in effective SNR are demonstrated. 
\end{abstract}

\section{Introduction and Motivation}
\vspace{-0.5ex}

Nonlinear distortions in signal propagation constitute a significant limitation for high-speed optical transmission systems. DSP algorithms have the potential to mitigate these impairments, but they suffer from high computational complexity. A common approach to alleviating this complexity is to design techniques based on simplified analytical expressions derived from the Manakov equation (ME), which is primarily used for the mathematical modeling of signal propagation in optical fibers. Interference cancellation using DSP after matched filtering is especially interesting in practice because it can be implemented with low complexity at 1 sample/symbol.

To design DSP algorithms for interference cancellation, a simple and precise channel model is essential. The ME has no closed-form solutions, and thus, it is commonly approximated using the framework of perturbation theory \cite{Vannucci2002b}. First-order regular perturbation (FRP) is widely used since it is well-suited for the performance assessment of fiber-optic transmission systems operating in the linear and the pseudo-linear regime. Furthermore, its time-domain form is well-suited to design coded modulation systems tailored to the fiber channel \cite{Bononi2020b}. Although FRP provides a reliable approximation, it can be computationally intensive and inaccurate at high transmission powers \cite{Bononi2020b, Barreiro2023}. Several works have aimed to reduce this complexity, with a growing interest in data-driven approaches \cite{Gao2019, Zhang2019, Melek2020, Barreiro2023}. 

A normalized batch gradient descent (NBGD) algorithm resulting in  \emph{optimized three-dimensional FRP coefficients}, also known as \textit{kernels}, was introduced in \cite{Barreiro2023}. It was shown in \cite{Barreiro2023} that FRP modeling via NBGD-optimized kernels reduces the complexity of the model and extends the range of powers for which FRP is valid. The so-called NBGD kernels were obtained using a data-driven approach, optimized per launch power, and make no simplifying assumptions, as often done in the literature. NBGD kernels are therefore interesting candidates to be at the core of future interference cancellation DSP algorithms. The use of NBGD kernels in such a setup has not been previously investigated, and thus, the problem of the potential advantages offed by NBGD kernels is open.

In this paper we study the potential theoretical gains offered by an interference cancellation scheme based on the NBGD kernels from \cite{Barreiro2023}. Our work follows \cite[Fig.~1(a)]{PortodaSilva2019}, where a genie-based interference cancellation architecture was considered. Such architecture gives a theoretical upper bound on the performance of schemes based on the newly introduced NBGD kernels. 
The results presented here show large potential gains in removing SPM in single-span single-channel systems. Our results also highlight the need for using NBGD kernels optimized for the specific launch power under consideration.

\section{System Model and Proposed NBGD-based Interference Cancellation}
\vspace{-0.5ex}

\begin{figure}[t]
    \centering
    \includegraphics[width=0.95\textwidth]{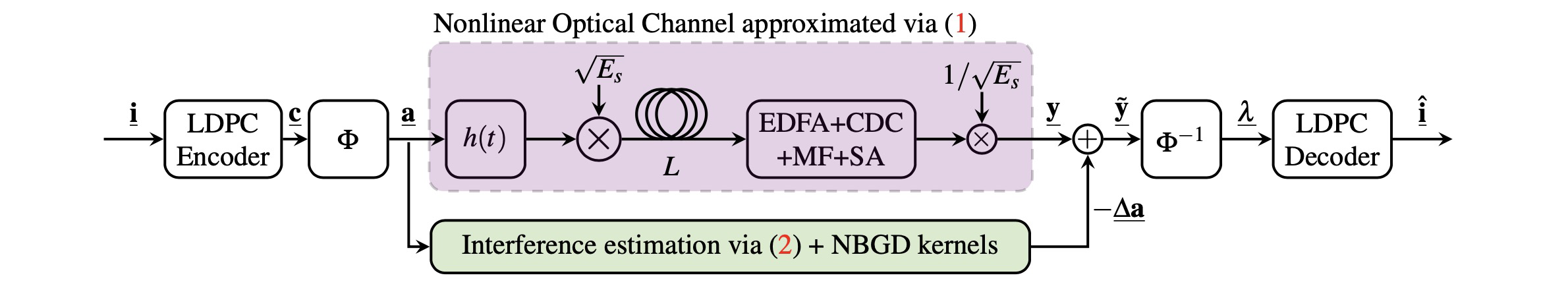}
    \vspace{-1ex}
    \caption{System model under consideration. The genie interference estimation uses NBGD kernels $S_{klm}$ and the transmitted sequence to ideally estimate the perturbative NLI term $\Delta \mathbf{a}$, setting an upper bound on performance. The traditional DSP chain is obtained setting $\Delta \mathbf{a}=0$.}
    \vspace{-5ex}
    \label{fig:system-model}
\end{figure}

Fig.~\ref{fig:system-model} shows the single-channel single-span fiber optical transmission system we consider, where a polarization-multiplexed signal propagates over a standard single mode fiber of length $L$. At the transmitter, the binary information sequence $\underline{\mathbf{i}}$ is encoded with a low-density parity-check (LDPC) code. A sequence of two-dimensional complex symbols $\underline{\mathbf{a}}=\ldots, \mathbf{a}_{n-1},\mathbf{a}_{n},$ $\mathbf{a}_{n+1},\ldots$ is then generated by a mapper $\Phi$. The constellation is normalized to unit energy. This sequence is linearly modulated using a pulse $h(t)$ with energy $E_s=P/2R_s$, to generate the transmitted signal, where $P$ and $R_s$ are the launch power and the symbol rate, respectively. Propagation is modeled by the attenuation-normalized ME \cite[Eq. (2)]{Barreiro2023}. After transmission over the fiber link, ideal coherent reception using an EDFA pre-amplifier that fully compensates fiber attenuation is followed by chromatic dispersion compensation (CDC), matched filtering (MF), sampling (SA) at the symbol rate, and scaling by $1/\sqrt{E_s}$. The nonlinear optical channel can then be seen as a channel with inputs $\underline{\mathbf{a}}$ and outputs ${\underline{\mathbf{y}}}$ (see Fig.~\ref{fig:system-model}).

 FRP modeling allows us to approximate the output of the nonlinear optical channel in Fig.~\ref{fig:system-model} via 
 \vspace{-1ex}
\begin{equation}\label{eq:approx}
{\mathbf{y}}_0 \approx {\mathbf{a}}_0 +\Delta \mathbf{a}_0 + {\mathbf{n}}_0,    
\vspace{-1ex}
\end{equation}
where ${\mathbf{a}}_0$ are the 2D-complex (column vector) transmitted symbols at time instant $k=0$ and ${\mathbf{n}}_0$ is the ASE noise, which is modeled as circularly symmetric Gaussian noise with zero mean and variance $\sigma_n^2$ per complex dimension. In FRP modeling, the nonlinear interference term $\Delta \mathbf{a}_0$ is given by
 \vspace{-1ex}
 \begin{equation}\label{eq:frp}
	 \Delta \mathbf{a}_0 = j\frac{8}{9} \gamma  E_s \sum_{(klm)\in \mathcal{M}}
  \left(\mathbf{a}_{k}^\dagger \mathbf{a}_{l}\right) \mathbf{a}_{m} 
  \hspace{0.1cm} S_{klm},
  \vspace{-1ex}
 \end{equation}
 where $j=\sqrt{-1}$, $\gamma$ is the fiber nonlinear coefficient, $S_{klm}$ are the perturbative coefficients or \emph{kernels}\cite{DarJLT2015}, and $(\cdot)^\dagger$ is conjugate transpose. In \eqref{eq:frp}, the triple summation is defined over the set $\mathcal{M}\triangleq \{(k,l,m)\in\mathbb{Z}^3 :-M\leq k,l,m \leq M\}$, where $|\mathcal{M}|=(2M+1)^3$, and $M$ denotes the {memory of the model}. Note that unlike many works in the literature, here we consider all the terms in the FRP approximation leading to a triple sum in \eqref{eq:frp}, which in turn requires three-dimensional kernels $S_{klm}$.

 \begin{wraptable}[]{r}{0.38\textwidth}
 \vspace{-5.5ex}
	\caption{Simulation parameters used in this paper}
 \vspace{-1ex}
	\centering
	{\footnotesize
		\begin{tabular}{c c}
			\toprule 
			Nonlinear parameter $\gamma$        & $1.2 \: \textrm{W}^{-1} \textrm{km}^{-1}$ \\
			Fiber attenuation $\alpha$          & $0.2 \: \textrm{dB/km}$ \\
			Group vel. disp. $\beta_2$           & $-21.7 \: \textrm{ps\textsuperscript{2}/km}$ \\
			Fiber length $L$                    & $230$\:km \\
			Symbol rate $R_s$                   & $60$\:Gbaud\\
			Pulse shape $h(t)$                  & RRC \\
			RRC roll-off factor                 & 0.01 \\
            Memory of the model $M$                 & 9\\
			\bottomrule
	\end{tabular}}
	\label{table:simparam} 
 \end{wraptable}
The kernels $S_{klm}$ can be computed numerically using their well-known integral form \cite[eq.~(4)]{DarJLT2015}. Such kernels, however, provide a good approximation of the NLI noise only in the weak nonlinearity regime. Optimized kernels $S_{klm}$ obtained by minimizing the root mean square error between a batch of true-transmission outputs and the output of the optimized model were proposed in \cite{Barreiro2023}. We call these kernels \textit{NBGD kernels}, which are known to provide a good approximation for the NLI even in the strongly nonlinear regime and when the model's memory $M$ is finite. We refer the reader to \cite[Sec.~IV]{Barreiro2023} for a detailed explanation of the optimization used.  
For the system model in Fig.~\ref{fig:system-model}, with parameters defined in Table~\ref{table:simparam}, it was found in \cite{Barreiro2023} that choosing $M=9$ is the best trade-off between accuracy and complexity. In this case, FRP enhanced with NBGD kernels, yields a 78\% complexity reduction compared to conventional FRP with $M=15$ while also extending the power range of validity of the FRP model above the optimum launch power\cite{Barreiro2023}. Thus, we use $M=9$ in this paper.

Fig.~\ref{fig:system-model} shows the proposed system that we use to quantify the potential gains offered by using the NBGD kernels for interference cancellation, which is identical to \cite[Fig.~1(a)]{PortodaSilva2019}. Unlike \cite{PortodaSilva2019}, where approximated kernels from \cite{Tao2014} were used (based on Gaussian pulses, high chromatic dispersion, low attenuation and the temporal phase/pulse-matching condition), here we consider optimized NBGD kernels without any assumptions/simplifications. We consider a genie approach, where the interference estimation block computes the NLI $\Delta \mathbf{a}_0$ using \eqref{eq:frp} and the transmitted symbols. In a practical scheme, this NLI estimation would be replaced by symbol estimates obtained from the FEC decoder or a decision feedback equalizer (DFE) for example, or would be the result of an iterative process between the demapper and the FEC decoder/DFE. An excellent summary of examples of such (iterative and noniterative) approaches is presented \cite[Sec.~I-A]{PortodaSilva2021}. Here we consider perfect knowledge of the transmitted symbols because we are interested in a theoretical upper bound on the performance of an interference cancellation based on NBGD kernels. 

As shown in Fig.~\ref{fig:system-model}, after the vector of NLI interference $\underline{\Delta \mathbf{a}}$ is estimated, the demapper is fed with a sequence of symbols $\tilde{\underline{\mathbf{y}}}=\underline{\mathbf{y}}-\underline{\Delta \mathbf{a}}$, which based on \eqref{eq:approx} should result in ${\mathbf{y}}_0 \approx {\mathbf{a}}_0 + {\mathbf{n}}_0$. This AWNG approximation will be valid as long as the approximation in \eqref{eq:approx} is valid, which given the results in \cite{Barreiro2023}, is expected to be precise even a few decibels above the optimum launch power. The last steps in the model in Fig.~\ref{fig:system-model} are LLR calculation ($\Phi^{-1}$) and FEC decoding. The LLR calculation is done assuming a bi-variate Gaussian distribution per polarization, where the mean vector and covariance matrix are estimated using a genie-aided approach. The LLRs are then used to compute the generalized mutual information (GMI) and also used for a soft-input LDPC decoder. 

\section{Numerical Results}
\vspace{-0.5ex}

\begin{figure}[!t]
    \centering
    \includegraphics[width=0.9\textwidth]{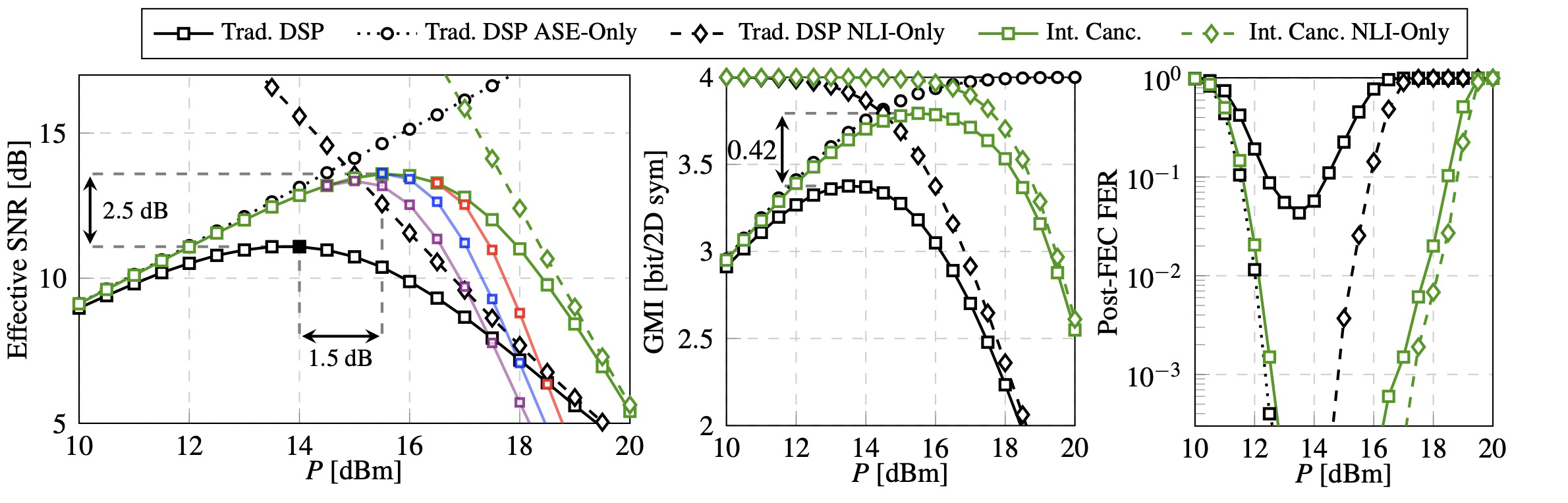}

    
    \vspace{-2ex}
    \caption{Effective SNR (left), GMI (middle), and post-FEC FER (right) for a traditional DSP chain (black lines) and the proposed (genie) interference cancellation scheme (green curves). The ASE-only case ($\gamma=0$) is shown with dotted lines and the NLI-only case (zero ASE noise) with dashed lines. Effective SNR results for mismatched interference cancellation are also shown (purple, blue, and red squares).}
    \vspace{-5ex}
    \label{fig:results}
\end{figure}

We consider a system with simulation parameters given in Table~\ref{table:simparam}. The modulation format is Gray-labeled PM-16QAM. An LDPC code from the 802.11 standard is used, with coding rate $3/4$ ($486$ information bits). The BP decoder uses a maximum of $40$ iterations. The results from the simulations are shown in Fig.~\ref{fig:results}. Three metrics are shown: effective SNR (left), GMI (middle), and post-FEC FER (right). The results for the traditional DSP chain (solid line with black squares) show the expected behavior: effective SNR (left) achieves a maximum at $P=14$~dBm, and then, performance decreases when $P$ is beyond the optimum launch power. Similar results are shown for GMI and post-FEC FER. Fig.~\ref{fig:results} also shows two other baselines: ASE noise only (dotted lines with black circles, $\gamma=0$) and NLI-only (dotted lines with black diamonds). In the former, as expected, a constant performance improvement is observed as $P$ increases. In the latter, ASE noise is removed and only NLI (and CD) is simulated. In this case, the performance of traditional DSP decreases as $P$ increases. This is again expected, as the system is then NLI-limited, which conventional DSP cannot deal with. At high input powers, the NLI-only case matches traditional DSP with ASE and NLI, as in that case, NLI dominates ASE.

The last set of results in Fig.~\ref{fig:results} are those related to the interference cancellation in Fig.~\ref{fig:system-model}. The effective SNR results (solid lines with green squares) show an increase in the optimum launch power of $1.5$~dB and an effective SNR increase of $2.5$~dB. The GMI results show an increase of $0.42$~bit/2D sym at optimum launch power. The SNR and GMI values decrease at very high power, which is due to the fact that the model can no longer predict accurately the symbols $\underline{\mathbf{y}}$ from split-step Fourier simulations. The post-FEC FER results in Fig.~\ref{fig:results} (right) also show that the minimum value obtained by the traditional DSP (black squares) is reduced from $4\cdot 10^{-2}$ to FER well below $10^{-4}$. A clear broadening of the ``bathtub'' results is also clearly visible. The results for the interference cancellation in the NLI regime only case (green lines with dashed diamonds in Fig.~\ref{fig:results}~(left)) show again that in the high-power regime, NLI dominate ASE noise, and thus, green dashed and green solid lines match. 

To conclude, we also performed a robustness analysis for the NBGD kernels. Fig.~\ref{fig:results}~(left) presents 3 solid lines with purple, blue, and red squares. These results show the effective SNR for \textit{mismatched NBGD kernels}, namely, when the NLI cancellation in Fig.~\ref{fig:system-model} is used with NBGD kernels that were optimized for a single launch power and then used for higher powers. The purple, blue, and red lines in Fig.~\ref{fig:results}~(left) show the NLI cancellation when kernels optimized for $14.5$~dBm, $15.5$~dBm (opt. launch power), and $16.5$~dBm, resp. These results show that NBGD kernels optimized for a single power \emph{cannot be used} for higher launch powers, which in turn suggests that the per-launch power optimization of kernels performed in \cite{Barreiro2023} is indeed needed.

\section{Conclusions}
\vspace{-0.5ex}

This paper showed large potential gains for nonlinearity cancellation using the recently introduced power-dependent NBGD kernels. Future work includes the design of a low-complexity receiver architecture that harvests most of the gains reported here, as well as comparisons to other cancellation schemes available in the literature based on simplified kernels.

\vspace{1ex}
\noindent\footnotesize{\textbf{Acknowledgements:} This work has received funding from the European Research Council (ERC) under the European Union's Horizon 2020 research and innovation programme (grant agreement No 757791).}
\vspace{-1ex}


\begin{thebibliography}{10}
\providecommand{\url}[1]{#1}
\csname url@samestyle\endcsname
\providecommand{\newblock}{\relax}
\providecommand{\bibinfo}[2]{#2}
\providecommand{\BIBentrySTDinterwordspacing}{\spaceskip=0pt\relax}
\providecommand{\BIBentryALTinterwordstretchfactor}{4}
\providecommand{\BIBentryALTinterwordspacing}{\spaceskip=\fontdimen2\font plus
\BIBentryALTinterwordstretchfactor\fontdimen3\font minus
  \fontdimen4\font\relax}
\providecommand{\BIBforeignlanguage}[2]{{%
\expandafter\ifx\csname l@#1\endcsname\relax
\typeout{** WARNING: IEEEtran.bst: No hyphenation pattern has been}%
\typeout{** loaded for the language `#1'. Using the pattern for}%
\typeout{** the default language instead.}%
\else
\language=\csname l@#1\endcsname
\fi
#2}}
\providecommand{\BIBdecl}{\relax}
\BIBdecl

\bibitem{Vannucci2002b}
A.~Vannucci et al., ``The {RP} method: a new tool for the
  iterative solution of the nonlinear {Schr{\"o}dinger} equation,'' JLT 2002.

\bibitem{Bononi2020b}
A.~Bononi et al., \emph{Fiber
  nonlinearity and optical system performance}, Springer Handbook of Optical Networks, Springer, 2020.

\bibitem{Barreiro2023}
A.~Barreiro et al., ``Data-driven enhancement of the
  time-domain first-order regular perturbation model,'' JLT, May 2023.

\bibitem{Gao2019}
Y.~Gao et al.,
  ``Reduced complexity nonlinearity compensation via principal component
  analysis...,'' OFC 2019.

\bibitem{Zhang2019}
S.~Zhang et al.,``{Field and lab experimental
  demonstration of nonlinear impairment compensation...},''
  \emph{Nature Communications} 2019.

\bibitem{Melek2020}
M.~M. Melek and D.~Yevick, ``{Nonlinearity mitigation with a perturbation based
  neural network receiver},'' \emph{Optical and Quantum Electronics}, Oct. 2020.


\bibitem{PortodaSilva2019}
E.~{Porto da Silva} et al.,``Perturbation-based {FEC}-assisted iterative nonlinearity
  compensation for {WDM} systems,'' JLT, Feb. 2019.

\bibitem{DarJLT2015}
R.~Dar et al.,``Inter-channel nonlinear
  interference noise in {WDM} systems: Modeling and mitigation,'' JLT, Mar. 2015.

\bibitem{Tao2014}
Z.~Tao et al., ``Analytical
  intrachannel nonlinear models to predict the nonlinear noise waveform,'' JLT, May
  2015.

\bibitem{PortodaSilva2021}
E.~{Porto da Silva} and M.~P. Yankov, ``Adaptive turbo equalization for
  nonlinearity compensation in {WDM} systems,'' JLT, Nov. 2021.

\end{thebibliography}


\end{document}